\documentclass[onecolumn,noshowpacs,nofootinbib,11pt]{revtex4}
\usepackage{float,xcolor,upgreek,yfonts,tikz}
\usepackage{color} 
\usepackage{textcomp,bm}
\usepackage{graphicx}
\usepackage{subfigure}
\usepackage{braket}
\usepackage{physics}

\def\b0{{\bf 0}}

\def\bdm{\begin{displaymath}}
\def\edm{\end{displaymath}}

\newcommand{\be}{\begin{eqnarray}}
\newcommand{\ee}{\end{eqnarray}}
\renewcommand{\b}{\hat b}

\newcommand{\Det}{\text{Det}\,}
\newcommand{\del}{\partial}

\newcommand{\source}{\varphi^{\scalebox{.65}{(0)}}}
\newcommand{\sourceG}{h^{\scalebox{.65}{(0)}}_{xy}}

\newcommand{\bmq}{{\vec{q}}}

\newcommand{\bmx}{{\vec{r}}}

\newcommand{\ikx}{ {i \omega t - i \bmq \cdot \bmx} }

\usepackage{graphicx,float}\usepackage[all]{xy}
\usepackage{amsmath,upgreek,accents}
\definecolor{applegreen}{rgb}{0.55, 0.71, 0.0}

\definecolor{amber}{rgb}{1.0, 0.55, 0.1}

\usepackage{amssymb}

\usepackage{textgreek}
\newcommand{\beq}{\begin{eqnarray}}

\newcommand{\eeq}{\end{eqnarray}}

\newcommand{\bea}{\begin{eqnarray}}
\newcommand{\eea}{\end{eqnarray}}

\usepackage[T3,T1]{fontenc}

\usepackage[colorlinks,hyperindex,unicode]{hyperref}
\definecolor{green1}{RGB}{0,128,0} 
\hypersetup{hidelinks,backref=true,pagebackref=true,hyperindex=true,colorlinks=true,breaklinks=true,urlcolor= blue}
\hypersetup{%
  colorlinks = true,
  linkcolor  = blue,
  citecolor = cyan,
}
\usepackage{bookmark,textgreek}
\usepackage{hyperref,color,xcolor}

\newcommand\orcidroldao{{\href{https://orcid.org/0000-0003-3978-532X}{\orcidicon}}}
\newcommand\orcidkuntz{{\href{https://orcid.org/0000-0003-2621-5715}{\orcidicon}}}
\newcommand{\orcidicon}{%
	\begin{tikzpicture}
	\draw[lime, fill=lime] (0,0)
		circle [radius=0.16]
		node[white] {{\fontfamily{qag}\selectfont \tiny ID}};
	\draw[white, fill=white] (-0.0625,0.095)
		circle [radius=0.007];
	\end{tikzpicture}	\hspace{-2mm}
}

\def\nn{\nonumber }

\begin{document}
\title{Transport and response coefficients in second-order dissipative relativistic hydrodynamics with quantum corrections: probing the quark-gluon plasma}

\author{Iberê Kuntz
\orcidkuntz\!\!
}
\affiliation{Departamento de F\'isica, Universidade Federal do Paran\'a, PO Box 19044, Curitiba -- PR, 81531-980, Brazil.}
\email{kuntz@fisica.ufpr.br}

\author{Roldao da Rocha
\orcidroldao\!\!
}
\affiliation{Federal University of ABC, Center of Mathematics,  Santo Andr\'e, 09210-580, Brazil.}
\email{roldao.rocha@ufabc.edu.br}

\begin{abstract}
A functional measure encompasses quantum  corrections 
and is explored in the fluid/gravity correspondence.
Corrections to response and transport coefficients in the $2^{\textsc{nd}}$-order dissipative relativistic hydrodynamics are proposed, including the ones to the pressure, the relaxation time, and the shear relaxation. 
Their dependence on the quark-gluon plasma (QGP) temperature 
sets a temperature dependence on the running parameter encoding the one-loop quantum gravity correction, driven by a functional measure. The experimental range of the bulk viscosity-to-entropy density ratio of the QGP 
%, obtained by five different analyses (JETSCAPE Bayesian model, Duke,  Jyväskylä-Helsinki-Munich, MIT-Utrecht-Gen\`eve, and Shanghai)  
corroborates the existence of the functional measure.
%Our results suggest that high-temperature plasmas could be used to test quantum gravity experimentally. 
\end{abstract}
\maketitle
\section{Introduction}
\label{s1}

AdS/CFT duality relates strongly correlated quantum  field theories with weakly coupled classical gravity. In the original setup, AdS/CFT states a well-defined relation between a 4-dimensional conformal field theory (CFT) and the geometry of a 5-dimensional anti-de Sitter (AdS) background describing gravity \cite{malda1,malda2,malda3}. 
Since classical dynamics of weakly-coupled gravity is described in a codimension one spacetime, AdS/CFT is inherently a holographic duality and the formal inception of other relevant gauge/gravity correspondences,    including AdS/QCD and AdS/CMT. 

When collective phenomena are studied in QCD and condensed matter,  one can observe strongly coupled quantum systems rearrange themselves in such a way that weakly-coupled degrees of freedom  dynamically emerge. Hence, a quantum system can be better formulated with respect to fields corresponding to the new emergent excitations. Gauge/gravity holographic dualities engender straightforward examples of this pattern, where the regarded emergent fields regulate gravity living in a codimension one spacetime. The additional dimension can be thought of as the energy scale of the dual quantum field theory.  Gauge/gravity dualities encompass the description of a variety of quantum systems, describing, for instance, the strong-coupled dynamics in QCD, black hole physics, quantum gravity,  relativistic hydrodynamics implementing the fluid/gravity correspondence, and holographic superconductors in condensed matter physics, among others \cite{Nastase:2015wjb}.

AdS/CFT can be analyzed in the long-wavelength limit, 
implementing a special type of gauge/gravity duality, known as the fluid/gravity correspondence. In this context, finite-temperature quantum systems, such as the quark-gluon plasma (QGP), can be studied using relativistic hydrodynamics. In this setup, the main features of viscous fluid flows can be 
specified by their transport and response coefficients, which enclose the relevant microscopic features of the relativistic hydrodynamics controlling the fluid flows. 
Dual to the hydrodynamics on the boundary,  gravity in the codimension one bulk can be characterized by an AdS black brane containing (at least) one non-degenerate event horizon. Therefore, the conformal field theory on the boundary, in the long-wavelength limit, is governed by the near-horizon regime of gravity in the bulk geometry \cite{Iqbal:2008by}.
AdS/CFT is thus a very advantageous instrument to calculate transport and response coefficients in relativistic hydrodynamics \cite{Son:2007vk}. 

Fluid/gravity correspondence establishes that  the energy-momentum tensor describing the strong-coupled boundary CFT is dual to the operators representing graviton fields in the AdS bulk. In the long-wavelength limit, the fact that the energy-momentum tensor is conserved leads to the relativistic hydrodynamical description of fluid flows. In this way, Einstein's equations in the AdS bulk can be related to Navier--Stokes equations on the boundary \cite{Bhattacharyya:2007vjd,Haack:2008cp,Policastro:2002se,Bernardo:2018cow,Bemfica:2017wps,Rocha:2022ind,Kovtun:2019hdm,Hoult:2020eho}. Several properties and features of relativistic hydrodynamics on the AdS boundary, describing a viscous fluid flow characterized by its response and transport  coefficients, have been scrutinized \cite{Policastro:2001yc}, also comprising soft-hair excitations \cite{Ferreira-Martins:2021cga,Ferreira-Martins:2019svk,Meert:2020sqv} and fermionic sectors reported in Refs. \cite{Meert:2018qzk,Bonora:2014dfa}.  Relativistic hydrodynamics is also a useful tool to study phase transitions and deconfinement in QCD \cite{DerradideSouza:2015kpt,Karapetyan:2018yhm,Karapetyan:2018oye,Karapetyan:2023kzs,Peralta-Ramos:2011fvn}. Ref. \cite{Gubser:2008yx} regarded black hole solutions with equations of state resembling the ones in QCD at zero chemical potential, and shear- and bulk-viscosities-to-entropy density ratios were calculated.
Refs. \cite{Critelli:2014kra,Ayala:2024jvc} studied anisotropic shear viscosity in  strongly-coupled plasmas with an external magnetic field, whereas momentum transport was scrutinized in this context in Ref. \cite{Finazzo:2016mhm}. Ref. \cite{Jena:2024cqs} investigated the QGP from the point of view of the Einstein--Born--Infeld-dilaton model, whereas other aspects of holography were investigated in this context  
\cite{Jena:2022nzw,Ballon-Bayona:2021tzw,Hippert:2023bel,Mamani:2022qnf}. Refs. \cite{Shen:2023awv}
 scrutinize the Bayesian analysis of the QGP bulk and shear viscosities at the high chemical potential region.

The use of relativistic viscous hydrodynamics to model the evolution of the QGP generated in ultrarelativistic heavy-ion collisions is currently ubiquitous. Important results using  
computational simulations comply with experimental data available from the Relativistic Heavy Ion Collider (RHIC) \cite{Gardim:2022yds}. 
One of the most straightforward, however successful, models describing heavy-ion collisions is the liquid drop model, whose evolution is governed by the equations of motion in relativistic
hydrodynamics. For them to be applied, the characteristic length scale of the system under scrutiny must be, in general, much larger than the mean free path ($\ell_{\textsc{mfp}}$). In this way, relativistic hydrodynamics consists of an effective theory capturing the low-frequency dynamics of wave modes with frequency and momentum much smaller than the inverse of $\ell_{\textsc{mfp}}$. Nuclear physics described by QCD has the mean free path of Fermi order, and the relativistic viscous  hydrodynamics works reasonably well to study collective phenomena in strongly interacting matter produced by the RHIC and the LHC
\cite{Czajka:2017bod}.

Transport and response coefficients in fluid dynamics measure how fast a perturbed system returns to equilibrium and are intrinsic tools underlying the hydrodynamical description. 
In the non-relativistic Navier--Stokes formulation, the dissipative currents, encompassing the heat flow, the bulk viscous pressure, and the shear energy-momentum tensor are supposed to have a linear dependence on the forces. They are indeed represented by the  gradients of the fluid 4-velocity, the temperature, and the 
chemical potential. The associated parameters of proportionality are the bulk viscosity,  the shear viscosity, and the heat conductivity. 
The Navier--Stokes description can be extended by considering
higher-order gradients of the 4-velocity, the temperature, and the chemical potential, leading to the bulk and shear relaxation times,  setting the characteristic time scales for dissipative currents to relax to their respective $1^{\textsc{st}}$-order solutions. 
The field-theoretical origin of the shear relaxation time was reported in Ref. \cite{Denicol:2011fa}, which showed the microscopic emergence of the shear relaxation time. Kubo's formula for the shear relaxation time was obtained for systems with conformal symmetry in Refs. \cite{Baier:2007ix,Moore:2010bu,Koide:2009sy}, calculating  the response of the referred systems to small perturbations for a background metric. Also, Kubo's  formula for the product of the bulk viscosity and the bulk relaxation time can be deduced from response functions studied in Ref. \cite{Hong:2010at}.

The calculation of the shear viscosity-to-entropy density ratio \cite{Kovtun:2004de} was extrapolated in Ref. \cite{Kuntz:2019omq} to include one-loop quantum gravitational corrections. Ref. \cite{Kuntz:2022kcw} also studied quantum gravity corrections due to the functional measure to some transport and response coefficients of the gauge theory. 
The functional measure is indeed required for the invariance of the effective action under field reparameterizations (and hence under gauge transformations) \cite{Vilkovisky:1984st,Mottola:1995sj,Fujikawa:1983im,Fujikawa:1979ay,Toms:1986sh,DeWitt:2003pm}. Geometrically, it can be introduced via the determinant of the configuration-space metric, following the same steps to obtain the integration measure in curved spacetimes. Unlike the gravitational analogy, however, the configuration-space metric is fixed from the onset. Thus, the classical action and the configuration-space metric must be given to fully specify a theory, both of which can be determined using effective field theory\footnote{One could promote the configuration-space metric to a dynamical object, but it brings back parameterization-dependence problems \cite{Casadio:2022ozp}.}.
The functional measure so obtained gives a correction to the effective action at one-loop order. In this paper, we build up on the work of Ref.~\cite{Kuntz:2022kcw} by computing the corrections due to the functional measure to other response and transport coefficients.

In the previous analysis of Ref.~\cite{Kuntz:2022kcw}, the calculation of the functional measure's running parameter $\upgamma$ was based on the comparison of the bulk viscosity-to-shear entropy ratio with experimental data of the QGP, including the JETSCAPE Bayesian model \cite{JETSCAPE:2020shq,JETSCAPE:2022cob},  the analysis by the Duke group \cite{Bernhard:2019bmu}, and the Jyväskylä-Helsinki-Munich group \cite{Parkkila:2021yha}. In this case, experimental lower and upper bounds on $\upgamma$ were found \cite{Kuntz:2022kcw}. In this paper, such analyses are updated and added to the analysis of the bulk viscosity-to-entropy density of the QGP by the MIT-Utrecht-Gen\`eve group \cite{Nijs:2020roc,Nijs:2022rme}, and the Shanghai group \cite{Yang:2022ixy} as well. Moreover, the analysis of other transport and response coefficients can propose to test quantum gravity in high-temperature scenarios, showing that the term due to the functional measure dominates as the temperature increases. A strong result in our current work makes the experimental range of the bulk viscosity-to-entropy density of the QGP, obtained by five different phenomenological analyses (JETSCAPE Bayesian model, Duke,  Jyväskylä-Helsinki-Munich, MIT-Utrecht-Gen\`eve, and Shanghai)  corroborate the existence of a non-vanishing renormalized parameter encoding the one-loop functional-measure quantum gravity correction.

This paper is organized as follows. In Sec. \ref{measure1}, we review the construction of the functional measure in AdS/CFT using effective field theory. In particular, we show how it leads to a correction to the classical action.
%Sec. \ref{s2} is dedicated to scrutinizing the functional measure on the gravity side, whereas 
The phenomenology of such correction on the gauge side of the AdS/CFT correspondence is then studied in Sec. \ref{s3} by describing the energy-momentum tensor as a perfect and viscous fluid. 
In Sec. \ref{s4}, we perform the $2^{\textsc{nd}}$-order hydrodynamics expansion to obtain further quantum gravity corrections to transport and response coefficients. We compare such results against numerical data from lattice simulations to constrain the functional measure correction.
In particular, the running  parameter that regulates quantum gravity corrections, due to the functional measure, is obtained as a function of the plasma temperature by analyzing the pressure, the relaxation time, and the shear relaxation in lattice QCD underlying the  QGP. \textcolor{black}{Sec. \ref{rhic} shows how to constrain the parameter responsible for carrying quantum gravity effects from experimental data of the QGP and heavy-ion collisions 
at LHC and RHIC. For it, the JETSCAPE Bayesian model will be used \cite{JETSCAPE:2020shq,JETSCAPE:2022cob}, together with the results by the Duke group \cite{Bernhard:2019bmu}, the Jyv\"askyl\"a-Helsinki-Munich group \cite{Parkkila:2021yha}, the MIT-Utrecht-Gen\`eve group  \cite{Nijs:2020roc,Nijs:2022rme}, and the Shanghai group \cite{Yang:2022ixy}. These analyses are thoroughly implemented in the context of a functional measure to bound the parameter that is responsible for quantum gravity effects.} 
We then draw our conclusions in Sec. \ref{s5}.

\section{Functional measure in AdS/CFT}
\label{measure1}

{\color{black}

The gauge/gravity correspondence can be succinctly stated via the GKP-Witten relation~\cite{malda3,malda2}:
\begin{equation}
    Z_{\scalebox{.65}{\textsc{gauge}}}
    =
    Z_{\scalebox{.65}{\textsc{gravity}}}
    \ .
    \label{gkp}
\end{equation}
Indeed, all quantities of interest in quantum field theory can be directly obtained from the generating functional\footnote{We shall adopt DeWitt's notation, where small-letter indices $i=(I,x)$ include both discrete indices $I$, denoted by capital letters, and the spacetime dependence $x$. Repeated small-letter indices amounts on summations over discrete indices and spacetime integrations.}:
\begin{equation}
	Z[J]
	=
	\int\mathrm{d}\mu[\varphi] e^{i \left( S[\varphi^i] + J_i \varphi^i \right)},
	\label{Z}
\end{equation}
where $S[\varphi^i]$ denotes the classical action for the underlying theory and $\varphi^i = \varphi^I(x)$ denotes arbitrary fields.  Eq.~\eqref{gkp} thus guarantees the equivalence of the corresponding theories.

As it is evident from Eq.~\eqref{Z}, such a duality strongly depends on the definition of the path integral, 
which, however, lacks a full-fledged mathematical construction, particularly regarding the integration measure.
From a geometrical point of view, a general configuration space requires the following definition of the integration measure:
\begin{equation}
    \mathrm{d}\mu[\varphi] = \mathcal{D}\varphi^i \sqrt{\Det G_{ij}}
    \ ,
    \label{measure}
\end{equation}
where $\mathcal{D}\varphi^i = \prod_i \mathrm{d}\varphi^i$ and $\Det G_{ij}$ denotes the functional determinant of the ultralocal configuration-space metric:
\begin{equation}
G_{ij} = G_{IJ}(\varphi) \, \delta(x,x')
\ .
\end{equation}
Ultralocality, namely the fact that $G_{ij}$ is proportional to the Dirac delta, is required by consistency with the local $S$-matrix theory. The functional determinant in this case evaluates to
\begin{equation}
	\ln \Det G_{ij}
	= 
	\delta^{(n)}(0) \int\mathrm{d}^nx \sqrt{-g} \tr\ln G_{IJ}
	\ .
	\label{detg}
\end{equation}
The term $\sqrt{\Det G_{ij}}$ is usually omitted in the path integral when dimensional regularization is used, in which case one writes $\delta^{(n)}(0) = 0$. However, the situation is not so straightforward  for such extreme divergences. Dimensional regularization hides the problem behind formal manipulations, not allowing for a careful analysis of the path integral measure. It might not even be applicable in these cases.

In the Wilsonian effective field theory, one regularizes the Dirac delta by implementing a cutoff. This can be done, for example, by using a Gaussian distribution, such that:
\begin{equation}
    \delta^{(n)}(0) = \frac{\Lambda^n}{(2\pi)^{n/2}}
    \ .
    \label{delta}
\end{equation}
With this regularization, Eq.~\eqref{detg} yields a non-trivial contribution from the configuration-space metric in \eqref{measure} to the effective action. Its contributions to some transport coefficients have already been calculated.

After introducing the non-trivial functional measure, one immediately faces the problem of determining the configuration-space metric. We stress that a choice of such a metric must be seen as part of the definition of the theory, along with the bare Lagrangian. Nonetheless, the Wilsonian approach comes again to our rescue. To leading order in the energy expansion, the most general configuration-space metric that satisfies all the required symmetries reads \cite{Casadio:2021rwj,Kuntz:2022tat}:
\begin{equation}
	\Det G_{ij}
	\sim
	\prod_x \det g_{\mu\nu}
	\ ,
	\label{detG2}
\end{equation}
whose precise coefficient depends on the fields present in the theory.
From Eqs.~\eqref{Z} and \eqref{detG2}, one then finds the contribution of the functional measure to the bare action:
\begin{equation}
	Z[J]
	=
	\int \mathcal{D}\varphi^i e^{i \left( S_{\scalebox{.45}{\textsc{eff}}}[\varphi^i] + J_i \varphi^i \right)},
\end{equation}
where\footnote{We used the freedom of choosing the normalizing factor $Z[0]$ to include the absolute value in the argument of the logarithm.}
\begin{equation}
    S_{\scalebox{.55}{\textsc{eff}}} = \int\mathrm{d}^nx \sqrt{-g}
    \left(
    \mathcal L
	-
    i \upgamma \, \tr\ln |g_{\mu\nu}|
	\right)
	\ ,
	\label{eq:newac}
\end{equation}
for some bare Lagrangian $\mathcal L$. Here, $\upgamma$ is the renormalized parameter, whose Wilsonian renormalization group equation reads
\begin{equation}
    \Lambda \frac{dZ}{d\Lambda} = 0
    \ .
    \label{RGE}
\end{equation}

One should note that the functional-measure correction in Eq.~\eqref{eq:newac} is of one-loop order. The usual one-loop correction $\ln\Det(S_{,ij})$ combines with the functional measure correction to wit
\begin{equation}
	\ln\Det(S^i_{\ ,j})
	=
	\ln\Det(G^{ik} S_{,kj})
	\ ,
\end{equation}
thus transforming the bilinear term $S_{,ij}$ into the linear operator $S^i_{\ ,j}$. Only the latter transforms covariantly under field redefinitions, which includes, in particular, spacetime diffeomorphisms. This stresses the fact that, despite the form of the correction in Eq.~\eqref{eq:newac}, there is no fundamental violation of the diffeomorphism invariance.
In fact, the functional measure \eqref{measure} transforms as a functional scalar density, canceling out the functional Jacobian from the transformation of $\mathcal{D}\varphi^i$. The net transformation thus leaves the path integral invariant under diffeomorphisms.

Because the usual $\ln\Det(S_{,ij})$ involves powers of curvatures and/or covariant derivatives, it is subdominant at low energies when compared to the derivative-free measure contribution. In leading order, one can then focus on the functional-measure correction, which is our main interest in this paper.

Note that each side of Eq.~\eqref{gkp} has its own functional measure. When using the AdS/CFT correspondence to perform calculations in the gauge theory, one has to take them both into account. However, it has been shown in Ref.~\cite{Kuntz:2022tat} that the functional measure from the gravity side does not lead to additional corrections for a diagonal spacetime metric. 
In this paper, we shall only be interested in
\textcolor{black}{the AdS$_5$-Schwarzschild background:
\begin{equation}
ds^2=-g_{tt}(u)dt^2+g_{uu}(u)du^2+g_{xx}(u)\left(dx^2+dy^2+dz^2\right),
\label{adsbh}
\end{equation}
with 
\begin{align}
g_{tt}(u)=\frac{L^2h(u)}{u^2},\qquad g_{uu}(u)=\frac{L^2}{u^2h(u)},\qquad\,g_{xx}(u)=\frac{L^2}{u^2},
\label{adss}
\end{align}
where $h(u) = 1 - u^4/u_0^4$ and $u_0$ is the event horizon\footnote{We have conveniently set all dimensionful parameters to unity.}}. Because Eq.~\eqref{adsbh} is diagonal, only the functional measure on the gauge side will yield non-trivial contributions, which we calculate in the next section.
We emphasize that the measures from each side of the correspondence are not dual to each other. In general (for non-diagonal metrics), calculations on the gauge theory would be affected by both measures, the one defined on the gauge theory itself and the one mapped from the gravity side.
}

\section{Response and transport coefficients and quantum corrections due to the functional measure}
\label{s3}

In quantum mechanics, perturbation theory sets in by considering a perturbation of a free Hamiltonian $
H=H_{\scalebox{.55}{\textsc{free}}}+\delta H(t),$ 
 with 
\begin{align}
\delta H(t) &= - \int d^3x\, {\scalebox{.9}{$\mathcal{O}$}}(\bmx) \source(t,\bmx). 
\label{eq:perturb}
\end{align}
External sources $\source$ encode the way how the operator ${\scalebox{.9}{$\mathcal{O}$}}$ responds to perturbations. The operator expected value can be expressed as 
$ 
\langle {\scalebox{.9}{$\mathcal{O}$}}(t,\bmx) \rangle = \text{tr}\,\rho(t) {\scalebox{.9}{$\mathcal{O}$}}(\bmx),$ 
where $\rho(t)$ stands for the density matrix associated with a canonical ensemble. When the source  $\source$ is turned on, the density matrix evolves, 
 and the perturbation of the operator can be written as \cite{Natsuume:2014sfa}
 \begin{align}
\delta \left\langle {\scalebox{.9}{$\mathcal{O}$}}(t, \bmx) \right\rangle 
=  
- i\!\int \!d^4r'  
\mathsf{G}_R^{{\scalebox{.65}{$\mathcal{O}$}}{\scalebox{.65}{$\mathcal{O}$}}}(t\!-\!\mathtt{t}, \bmx\!-\!\bmx{\,'}) \source(\mathtt{t},\bmx{\,'}),
\label{eq:linear_response}
\end{align}
where the retarded Green's response function reads
\beq
\mathsf{G}_R^{{\scalebox{.65}{$\mathcal{O}$}}{\scalebox{.65}{$\mathcal{O}$}}}(t-\mathtt{t}, \bmx-\bmx{\,'}) = -i \theta(t-\mathtt{t})
\left\langle [{\scalebox{.9}{$\mathcal{O}$}}(t,\bmx), {\scalebox{.9}{$\mathcal{O}$}}(\mathtt{t},\bmx{\,'}) ] \right\rangle,
\label{grr}
\eeq
whose Fourier transformation can be written as 
$ 
\delta \langle {\scalebox{.9}{$\mathcal{O}$}}(q) \rangle = - \,\mathsf{G}_R^{{\scalebox{.65}{$\mathcal{O}$}}{\scalebox{.65}{$\mathcal{O}$}}}(q)\source(q)$, 
where $q_\mu = (\upomega, \bmq)$. The response function can be therefore written as \be
\mathsf{G}_R^{{\scalebox{.65}{$\mathcal{O}$}}{\scalebox{.65}{$\mathcal{O}$}}}(q) = -i \!\int\! d^4r\, e^\ikx 
\theta(t)  \left\langle \left[{\scalebox{.9}{$\mathcal{O}$}}(t,\bmx), {\scalebox{.9}{$\mathcal{O}$}}(0,\vec{0}) \right] \right\rangle.\label{grf}
\ee
Eq. (\ref{grf}) can be forthwith computed by AdS/CFT tools. 
In the context of fluid/gravity correspondence, any given Lagrangian can be perturbed as 
\beq
    \delta{\cal L} = h_{\mu\nu}(t) T^{\mu\nu}(\vec{r}),\label{pl}
\eeq
where the perturbation of the energy-momentum tensor is read off the 2-point retarded Green function acting on the metric perturbation, as
\beq
    \delta \langle T^{\mu\nu} \rangle &=& -\mathsf{G}_R^{\mu\nu,\rho\sigma} h_{\rho\sigma}(t), 
    \label{eq:response_emtensor} 
    \eeq
    where 
    \beq
        \mathsf{G}_R^{\mu\nu,\rho\sigma} &=& -i \int d^4r\, e^\ikx 
    \theta(t)  \left\langle [T^{\mu\nu}(t,\bmx), T^{\rho\sigma}(0,\vec{0}) ] \right\rangle~.\label{rgf}
\eeq
The energy-momentum tensor couples to metric perturbations on the boundary gauge theory. For the action \eqref{eq:newac}, the effective energy-momentum tensor can be expressed by 
\begin{align}
    T_{\mu\nu}^{\scalebox{.59}{\textsc{eff}}}
    &=
    \frac{2}{\sqrt{-g^{\scalebox{.65}{(0)}}}}
    \left[
        \frac{\partial\left(\sqrt{-g^{\scalebox{.65}{(0)}}} \mathcal L_{\scalebox{.59}{\textsc{eff}}}\right)}{{\scalebox{.98}{$\partial$}} {g^{\scalebox{.65}{(0)}}}^{\mu\nu}}
        - {\scalebox{.98}{$\partial_\rho$}}\frac{{\scalebox{.98}{$\partial$}}\left(\sqrt{-g^{\scalebox{.65}{(0)}}} \mathcal L_{\scalebox{.59}{\textsc{eff}}}\right)}{{\scalebox{.98}{$\partial$}}\left({\scalebox{.98}{$\partial_\rho$}} {g^{\scalebox{.65}{(0)}}}^{\mu\nu}\right)}
    \right]\nonumber
    \\
    &=
    T_{\mu\nu}
    + i \upgamma \left(2 +  \tr\ln \left|g_{\rho\sigma}^{\scalebox{.65}{(0)}}\right| \right) g_{\mu\nu}^{\scalebox{.65}{(0)}}
    \ ,
    \label{eq:effT}
\end{align}
carrying an extra term that encodes quantum corrections due to the functional measure. The arbitrary metric $g_{\mu\nu}^{\scalebox{.65}{(0)}}$ has been considered arbitrary heretofore. From now on, off-diagonal perturbations of a flat space approach will be taken into account.

\subsection{Perfect fluid flows in hydrodynamics}

Hydrodynamic fluid flows can be described by the energy-momentum tensor containing quantum corrections, with the conservation law 
\begin{equation}
    \nabla^\mu T^{\scalebox{.59}{\textsc{eff}}}_{\mu\nu}
    =
    0
\end{equation}
compatible with the invariance of the path integral under diffeomorphisms.
In this scenario, perfect fluids are governed by the following constitutive relation:
\be
\left(T^{\scalebox{.59}{\textsc{eff}}}\right)^{\mu\nu} = \left(\upepsilon^{\scalebox{.59}{\textsc{eff}}}+P^{\scalebox{.59}{\textsc{eff}}}\right)u^\mu u^\nu + P^{\scalebox{.59}{\textsc{eff}}} {g^{\scalebox{.65}{(0)}}}^{\mu\nu},
\label{eq:EM_constitutive}
\ee
where $u^\mu(x)$ denotes the 4-velocity and $P^{\scalebox{.59}{\textsc{eff}}}$  stands for the effective pressure, whereas $\upepsilon^{\scalebox{.59}{\textsc{eff}}}$ is the fluid energy density.
The quantum corrections carried by the effective energy-momentum tensor in  Eq.~\eqref{eq:effT} induce Eq. \eqref{eq:EM_constitutive} to yield  \cite{Kuntz:2022kcw}
\begin{align}
    \upepsilon^{\scalebox{.59}{\textsc{eff}}}
    &=
    \upepsilon
    - 2 i \upgamma
    \ ,
    \label{em2}
    \\
    P^{\scalebox{.59}{\textsc{eff}}}
    &=
    P + 2i\upgamma
    \ ,
    \label{peff}
\end{align}
where $\upepsilon$ and $P$ denote, respectively, the energy density and the pressure in relativistic hydrodynamics without quantum corrections due to the functional measure. The imaginary part of  Eq. \eqref{em2} points to the instability of degrees of freedom in the fluid  and measures its lifetime. 
 The magnitude of this instability is driven by the running parameter $\upgamma$, whose dependence on the temperature will be analyzed for a non-conformal strongly-coupled plasma, for the probe limit and $g_{\scalebox{.65}{\textsc{YM}}}\gg 1$,  $g_{\scalebox{.65}{\textsc{YM}}}^2N_c\gg 1$. 
 
\subsection{Viscous fluid flows}
When viscous fluids are taken into account, the constitutive equation \eqref{eq:EM_constitutive} must be replaced by a generalization that includes 1$^\textsc{st}$-order derivatives of the 4-velocity:
\begin{align}
\!\!\!\!(T^{\scalebox{.59}{\textsc{eff}}})^{\mu\nu} &= \left(\upepsilon^{\scalebox{.59}{\textsc{eff}}}\!+\!P^{\scalebox{.59}{\textsc{eff}}}\right)u^\mu u^\nu \!+\! g^{(0)\mu\nu} 
 P^{\scalebox{.59}{\textsc{eff}}} \!-\! \Uppi^{\mu\rho} \Uppi^{\nu\sigma} \left[
\eta^{\scalebox{.59}{\textsc{eff}}}\! \left(\nabla_{(\rho} u_{\sigma)}  
\!-\! \frac{2}{3} g^{\scalebox{.65}{(0)}}_{\rho\sigma} \nabla \cdot  \vec{u} \right) 
\!+\! \upzeta^{\scalebox{.59}{\textsc{eff}}} g^{\scalebox{.65}{(0)}}_{\rho\sigma} \nabla \cdot  \vec{u} \right],
\label{eq:dissipative_curved}
\end{align}
where 
 $\Uppi^{\mu\nu} := g^{{\scalebox{.65}{(0)}}\mu\nu} + u^\mu u^\nu$ denotes the projection on spatial directions. We denote by $\upzeta^{\scalebox{.59}{\textsc{eff}}}$ and $\eta^{\scalebox{.59}{\textsc{eff}}}$ the effective bulk and shear viscosities, respectively.  The bulk viscosity measures the mean free path concerning any  process where the particle
number is not conserved. The shear viscosity appears in the response of the energy-momentum tensor to a small and slow-varying metric perturbation,
\be
g^{\scalebox{.65}{(0)}}_{\mu\nu}dx^\mu dx^\nu = \bar{g}_{\mu\nu}dx^\mu dx^\nu + 2h^{\scalebox{.65}{(0)}}_{xy}(t)dxdy.
\label{eq:metric_kubo}
\ee
The perturbation of the dissipative term 
\beq
\uptau^{xy} := - \Uppi^{x\rho} \Uppi^{y\sigma} \left[
\eta^{\scalebox{.59}{\textsc{eff}}}\! \left(\nabla_{(\rho} u_{\sigma)}  
- \frac{2}{3} g^{\scalebox{.65}{(0)}}_{\rho\sigma} \nabla \cdot  \vec{u} \right) 
+ \upzeta^{\scalebox{.59}{\textsc{eff}}} g^{\scalebox{.65}{(0)}}_{\rho\sigma} \nabla \cdot  \vec{u} \right]
\eeq in Eq. \eqref{eq:dissipative_curved} can be evaluated at linear order. From the fact that $
\nabla_x u_y 
= \frac{1}{2} \del_t h^{\scalebox{.65}{(0)}}_{xy}$ and that the gradient term $\nabla \cdot \vec{u}$ carries $2^{\textsc{nd}}$-order terms in the $h_{\mu\nu}$ perturbation, it implies that  
\beq
\delta \langle \uptau^{xy}  \rangle 
= - \eta^{\scalebox{.59}{\textsc{eff}}} \del_t h^{\scalebox{.65}{(0)}}_{xy},
\eeq
whose Fourier transformation can be written as 
\be
{
\lim_{q\to0}\delta \langle \uptau^{xy}(\upomega,\bmq)  \rangle = i\upomega \eta^{\scalebox{.59}{\textsc{eff}}} \sourceG~.
}
\label{eq:Txy_vs_eta}
\ee
When Eqs. \eqref{eq:response_emtensor} and \eqref{eq:Txy_vs_eta} are compared one arrives at the Kubo's formula 
\be
\eta^{\scalebox{.59}{\textsc{eff}}} = - \lim_{\substack{\upomega\rightarrow 0\\
 q\to0}} \frac{1}{\upomega} \Im\, \mathsf{G}_R^{xy,xy}(\upomega,\bmq)~.
\label{eq:viscosity_kubo}
\ee
Analogously, the retarded Green's response function in Eq. (\ref{rgf}) gives the bulk viscosity \cite{Gubser:2008sz,Buchel:2007mf,Czajka:2017bod}:
\beq
\upzeta^{\scalebox{.59}{\textsc{eff}}} =\lim_{\substack{{\upomega}\rightarrow 0\\
 q\to0}} \frac{1}{\upomega}\Im \mathsf{G}_R^{PP}(\upomega, \vec{q}),\label{eq:viscosity_kubo11}
 \eeq
 where
 \beq
 \mathsf{G}_R^{PP}(\upomega, \vec{q}) &=& \frac{k_ik_jk_mk_n}{k^4}\left[\mathsf{G}_R^{ij,mn}(\upomega, \vec{q})+\frac13 \delta_{ab}T^{ab}\left(\delta^{im}\delta^{jn}+\delta^{in}\delta^{jm}-\delta^{ij}\delta^{mn} \right)\right]+\frac13 \delta_{ij}T^{ij}\nonumber\\
 &&  -\frac43 \mathsf{G}_R^{xy,xy}(\upomega, \vec{q})\label{zetad}
 \eeq
is the response function to longitudinal fluctuations. 
When the perturbation \eqref{eq:metric_kubo} is implemented in Eq.~\eqref{eq:effT}, it yields corrections of the energy-momentum tensor due to the functional measure, resulting in the effective energy-momentum tensor given by  \begin{equation}
    T^{\scalebox{.59}{\textsc{eff}}}_{\mu\nu}
    =
    T_{\mu\nu}
    + 2i\upgamma \eta_{\mu\nu}
    \ .
    \label{finalT}
\end{equation}
Now, one can read off the effective bulk and shear viscosities from Eqs.~(\ref{eq:viscosity_kubo}, \ref{eq:viscosity_kubo11}, \ref{finalT}), as
\begin{align}   
    \upzeta^{\scalebox{.59}{\textsc{eff}}}
    &=\upzeta+4i\upgamma,\label{bul0}\\
     \eta^{\scalebox{.59}{\textsc{eff}}}
    &=
    \eta
\,.\label{bul}
\end{align}
The effective shear viscosity does not carry any quantum gravity corrections in the setup considering the functional measure, at least when flat-space backgrounds are regarded. However, the effective bulk viscosity does carry quantum gravity corrections \cite{Kuntz:2022kcw}. Eq.~\eqref{eq:effT} evinces no correction for diagonal metrics and solely  curved background metrics with non-diagonal terms induce quantum gravity corrections to the shear viscosity. The entropy density at leading order also carries no quantum corrections, since they cancel out the ones on the effective pressure, 
\beq\label{seffs}
s^{\scalebox{.59}{\textsc{eff}}} = \frac{P^{\scalebox{.59}{\textsc{eff}}}+\upepsilon^{\scalebox{.59}{\textsc{eff}}}}{T} = s.
\eeq
It yields the Kovtun--Son--Starinets result  \cite{Kovtun:2004de} to be invariant in the functional measure approach:
\beq
\frac{\eta^{\scalebox{.59}{\textsc{eff}}}}{s^{\scalebox{.59}{\textsc{eff}}}}
=
\frac{\eta}{s}.
\eeq

\section{$2^{\textsc{nd}}$-order derivative expansion of relativistic hydrodynamics}
\label{s4}

Transport and response coefficients, in the $2^{\textsc{nd}}$-order hydrodynamical formulation, can carry significant signatures of quantum gravity, arising from the presence of a functional measure.
 One can calculate these coefficients in the ${\cal N}=4$ super Yang--Mills dual plasma \cite{Arnold:2011ja,Buchel:2016cbj,Kovtun:2011np,Finazzo:2014cna}. 
$2^{\textsc{nd}}$-order hydrodynamics provides relevant tools to probe nuclear collisions, in the ultrarelativistic limit making the effective energy density significant to deconfinement of hadronic matter into the QGP  \cite{Schenke:2012wb,GoncalvesdaSilva:2017bvk,Oliveira:2020yac,Braga:2022yfe,Romatschke:2009kr,daRocha:2024lev,Casadio:2016zhu,Abdalla:2009pg,Rougemont:2023gfz,Braga:2023qej}. 

In the $2^{\textsc{nd}}$-order hydrodynamics setup, the dissipative part of the energy-momentum tensor
reads \cite{Baier:2007ix} 
\beq\label{gr1}
\Uppsi^{\mu\nu} &=& -\eta \upsigma^{\mu\nu}- \eta\tau_\pi \left[  u_\rho\nabla^\rho {}^\llcorner\upsigma^{\mu\nu}{}^\lrcorner 
 + \frac 12 \upsigma^{\mu\nu}
    (\nabla \cdot \vec{u}) \right]  
  + \upkappa\left[R^{\llcorner\mu\nu\lrcorner}-2 u_\alpha R^{\alpha\llcorner\mu\nu\lrcorner\beta} 
      u_\beta\right]\nonumber\\
  &&+ \frac{\uplambda_1}{\eta^2} {\upsigma^{\llcorner\mu}}_\lambda \upsigma^{\nu\lrcorner\lambda}
  - \frac{\uplambda_2}{\eta} {(\Uppsi)^{\llcorner\mu}}_\lambda \Omega^{\nu\lrcorner\lambda}
  + \uplambda_3 {\Omega^{\llcorner\mu}}_\lambda \Omega^{\nu\lrcorner\lambda}+2\kappa^{*} \,u_\rho u_\sigma {R}^{\rho \lrcorner \mu\nu \llcorner \sigma} \nonumber\\
  &&-3\eta\tau_\pi \left(\frac{1}{3}-c_s^2\right) \, \sigma^{\mu\nu}\,\frac{1}{3}\nabla_\rho u^\rho+\lambda_4 (\nabla^{\lrcorner\mu}\ln s)(\nabla^{\nu\llcorner} \ln s)\,,
\eeq
involving the Riemann and Ricci tensors, and the  traceless transverse tensor
$
  \upsigma^{\mu\nu} = 2{}^\lrcorner\nabla^{\mu} u^{\nu}{}^\llcorner$, 
where the notation 
\begin{equation}
  {}^\lrcorner A^{\mu\nu}{}^\llcorner
 = - \frac1{3} \Uppi^{\mu\nu} \Uppi^{\alpha\beta} A_{\alpha\beta}
+ \frac12 \Uppi^{\mu\alpha} \Uppi^{\nu\beta}
     A_{(\alpha\beta)} 
 \end{equation} 
for an arbitrary $2^{\textsc{nd}}$-rank tensor $A^{\mu\nu}$ is adopted. The symmetrization symbol $A_{(\alpha\beta)}=\frac{1}{2!}\left(A_{\alpha\beta}-A_{\beta\alpha}\right)$ is also implicit. The vorticity is given by 
\begin{equation}
  \Omega^{\mu\nu} = 
  \frac12   
    \Uppi^{\mu\theta}\Uppi^{\nu\tau}
    \left(\nabla_{\theta} u_{\tau}-\nabla_{\tau} u_{\theta}\right)\,.
\end{equation}
The  $\uplambda_1$, $\uplambda_2$, $\uplambda_3$, and $\uplambda_4$ parameters in Eq. (\ref{gr1}) are transport coefficients in $2^{\textsc{nd}}$-order hydrodynamics   \cite{Romatschke:2009kr}. 
%and the expression $u_\mu\nabla^\mu\eta = -\eta\,\nabla \cdot \vec{u}$ was used. 
The response coefficient 
\beq\label{kk}
\upkappa =\lim_{\substack{{q}\to 0\\\upomega\to 0}}\frac{\partial^2}{\partial{q^2}} \mathsf{G}^{xy,xy}_R(\upomega,\vec{q})
\eeq is the gravitational susceptibility of a QGP and 
enters the expression of the relaxation time for the shear relaxation  of  fluids with viscosity, which can be written as \cite{Baier:2007ix,Finazzo:2014cna,Natsuume:2007ty,Rocha:2022fqz,Rocha:2021zcw}. 
 \beq
 \tau_\pi=\frac{1}{2\eta}\left(\lim_{\substack{{q}\to 0\\\upomega\to 0}}\frac{\partial^2}{\partial{\upomega^2}} \mathsf{G}^{xy,xy}_R(\upomega,\vec{q})-\upkappa+T\frac{d\upkappa}{dT}\right).\label{tt}
\eeq
As a thermodynamic coefficient, $\upkappa$ can be computed from lattice QCD \cite{Buchel:2022fqq}.

One can derive Kubo's formul\ae\, for the $2^{\textsc{nd}}$-order hydrodynamics coefficients, considering a 
uniform system in equilibrium at an initial state in Minkowski flat space,  introducing 
perturbatively weak and slowly-varying
 nonuniformity.  Again writing the metric as the sum of the Minkowski metric and a perturbation, 
$h_{\mu\nu}(x)$, the perturbation metric tensor couples to the
energy-momentum tensor $T^{\mu\nu}$ as, e.g., in Eq. (\ref{pl}), generating the expansion in correlation
functions of multiple energy-momentum tensors. The associated  coefficients in this expansion consist of the response
of the energy-momentum to fluid nonuniformity. 
One can regard the expectation value
$\langle T^{\mu\nu}(0) \rangle$ for a system in
equilibrium in an initial time $t_0$,  at temperature $T$ and consider the  metric
perturbation $h_{\mu\nu}(x)$, with $h_{\mu\nu}(t)\equiv 0$ for all $t \leq t_0$.  The
energy-momentum tensor reads 
\begin{eqnarray}
\langle T^{\mu\nu}(0) \rangle & = & {\rm Tr}\: e^{-\beta H}
 {\rm \bar{\textsc{T}}\,\textsc{exp}}\left( \int_{t_0}^0  dt' i H[h(t')] \right)
\left(T^{\mu\nu}\right) {\textsc{T\,exp}}\left(  \int_{t_0}^0 dt'' (-i) H[h(t'')] \right)
\end{eqnarray}
where  $\bar{\textsc{T}}\,\textsc{exp}$ [${\textsc{Texp}}$] is the anti-time [time]-ordered exponential operators, $H[h(t)]$ is the metric-dependent Hamiltonian,
and $\beta$ is the inverse of the temperature \cite{Wang:1998wg,Wang:1998wg1,Chou:1984es}.
Independent metric perturbations for the
${\textsc{T}}$-ordered and $\bar{\textsc{T}}$-ordered evolution operators can be introduced to construct the generating functional
\begin{eqnarray} \!\!\!\!\!
W[h_1,h_2] & \equiv & \ln {\rm Tr}\: e^{-\beta H}
{\rm \bar{\textsc{T}}\textsc{exp}}\left(i \int_{t_0}^\infty dt' H[h_2(t')] \right) {\textsc{T\,exp}} \left( -i \int_{t_0}^\infty dt' H[h_1(t')] \right).
\end{eqnarray}
One then defines the average metric perturbation
\beq
h_{\scalebox{.59}{\textsc{av}}} = \frac12(h_1+h_2)\eeq and the average energy-momentum tensor
$T_{\scalebox{.59}{\textsc{av}}} = \frac12({T_1+T_2})$, as well as the variables
$h_a = h_1-h_2$, $T_a = T_1-T_2$.  Variation with respect
to $h_a$ yields 
\begin{equation}
\langle T^{\mu\nu}_{\scalebox{.59}{\textsc{av}}}(x) \rangle = \frac{-2i}{\sqrt{-g}} \frac{\partial W}{\partial (h_{a})_{\mu\nu}(x)}.
\end{equation}
 After taking the derivative with respect to $h_a$, one can set it to zero, implying that $h_{\scalebox{.59}{\textsc{av}}} = h$, as the 
background $h_1=h_2=h_{\scalebox{.59}{\textsc{av}}}=h$ remains the case of interest, although considering the term $h_a\neq0$ encodes 
quantum fluctuations in the metric. 
One therefore obtains 
\begin{eqnarray}
\label{Texpand}
\!\!\!\!\!\!\!\!\!\!\!\!\langle T_{\scalebox{.59}{\textsc{av}}}^{\mu\nu}\rangle_{h} & \!=\! & \langle T^{\mu\nu}\rangle_{h=0}
\!-\! \frac{1}{2} \int d^4 x \mathsf{G}_{ra}^{\mu\nu,\sigma\rho}(0,x) h_{\sigma\rho}(x) \!+\! \frac{1}{8} \int d^4 x d^4 y \,
 \mathsf{G}_{raa}^{\mu\nu,\sigma\rho,\tau\alpha}(0,x,y)
        h_{\sigma\rho}(x) h_{\tau\alpha}(y),
\end{eqnarray}
up to terms in ${\cal O}(h_{\mu\nu}^3)$, for 
 the
retarded correlation function reading \cite{Wang:1998wg,Moore:2010bu,Romatschke:2009kr}
\begin{eqnarray} \hspace{-0.2em}
\mathsf{G}^{\mu\nu,\alpha\beta,\ldots}_{s a \ldots}(0,x,\ldots)  = \lim_{{g_{\mu\nu}\to \eta_{\mu\nu}}}
 \frac{(-1)^{n-1}2^ni \;\partial^n W}{\partial (g_{a})_{\mu\nu}(0)
         \partial (g_{r})_{\alpha\beta}(x) \ldots }
       = (-i)^{n-1}
\left\langle T^{\mu\nu}_{\scalebox{.59}{\textsc{av}}}(0) T^{\alpha\beta}_a(x) \ldots
     \right\rangle.
\end{eqnarray}
  Eqs. (\ref{kk}, \ref{tt}) hold for both conformal and non-conformal fluids. The energy-momentum tensor, in momentum space, reads \cite{Arnold:2011ja}
\bea
\!\!\!\!\!\!\!\!\langle T^{\mu\nu}({q})\rangle_h&=&\langle T^{\mu\nu}\rangle_{h=0}
-\frac{1}{2} \int d^4{q}_1 
\delta^4({q}-{q}_1) \mathsf{G}^{\mu\nu,\sigma\rho}({q};-{q}_1)h_{\sigma\rho}({q}_1)\nn\\
&&+ \frac{1}{8} \int d^4{q}_1 \int d^4{q}_2 
\delta^4({q}-{q}_1-{q}_2) \mathsf{G}^{\mu\nu,\sigma\rho,\tau\alpha}({q};-{q}_1,-{q}_2)h_{\sigma\rho}({q}_1)h_{\tau\alpha}({q}_2)+\dots \label{npo FT}
\eea 
Assuming that $
q_1^\mu=(\upomega_1,0,0,\mathtt{q}_1)$ and $ q_2^\mu=(\upomega_2,0,0,\mathtt{q}_2)$, therefore, the transport coefficients  $\uplambda_1,\uplambda_2$, and $\uplambda_3$ can be read off from 3-point  correlators, in the strong-coupling regime, when evaluating the cubic Witten diagrams, as \cite{Arnold:2011ja,Bhattacharyya:2007vjd}
\beq
&&\lim_{\substack{\upomega_1\to0 \\ \upomega_2\to 0}}
\frac{\partial}{\partial{\upomega_2}}\frac{\partial}{\partial{\upomega_1}} \lim_{\substack{\mathtt{q}_2\to0 \\ \mathtt{q}_1\to 0}}\mathsf{G}^{xy,zx,yz}=\eta\tau_\pi-\uplambda_1,\nn\\
&&\lim_{\substack{\upomega_1\to 0\\\mathtt{q}_2\to 0}}\frac{\partial}{\partial{\upomega_1}}\frac{\partial}{\partial{\mathtt{q}_2}}\lim_{\substack{\mathtt{q}_2\to 0\\\upomega_1\to 0}}\mathsf{G}^{xy,yz,x0}=\frac 12\eta\tau_{\Pi}-\frac 14\uplambda_2,\nn\\
&&\lim_{\substack{\mathtt{q}_1\to0\\\mathtt{q}_2\to0}}
     \frac{\partial}{\partial{\mathtt{q}_2}} \frac{\partial}{\partial{\mathtt{q}_1}}
      \lim_{\substack{\upomega_1\to0\\\upomega_2\to0}}
            \mathsf{G}^{xy,x0,y0}
   =
   -\frac14 \uplambda_3,\label{preview}
\eeq
where $\tau_\Pi$ is the bulk relaxation. 
Throughout this work after Eq. (\ref{adss}) we put all dimensionful parameters to unity. For completeness, we will restore the event horizon $u_0$ hereon. 
Let us first discuss the coefficient $\upkappa$ in Eq. \eqref{kk}, by defining the regularized quantity \begin{align}
\upkappa_\varepsilon := -\lim_{\substack{\mathtt{q}\to0\\\upomega\to0}}
\frac{\partial^2}{\partial q^2}\textsc{F}(\upomega,q;\varepsilon)
=\frac{1}{2\varepsilon^2}+\int_{u_0}^\varepsilon du\sqrt{-g^{\scalebox{.65}{(0)}}}g^{xx}(u).
\label{3.1}
\end{align}
 Also, the expression that follows lacks any ultraviolet divergence  \cite{Finazzo:2014cna}:
\begin{align}
\upkappa(T)=& \lim_{\varepsilon\rightarrow 0}\left(\upkappa_\varepsilon(T) -\upkappa_\varepsilon(T_{\scalebox{.55}{{\textsc{high}}}})\right) +\frac{T^2_{\scalebox{.55}{{\textsc{high}}}}N_c^2}{8}-\upkappa_0& \nonumber\\
\approx &\lim_{\varepsilon\rightarrow 0}\left[\int_{u_0(T)}^\varepsilon du\,\sqrt{-g^{\scalebox{.65}{(0)}}} g^{xx}(u)\biggr|_{u_0(T)} -\int_{u_{\scalebox{.55}{{\textsc{high}}}}}^\varepsilon du\,\sqrt{-g^{\scalebox{.65}{(0)}}} g^{xx}(u)\biggr|_{u_{\scalebox{.55}{{\textsc{high}}}}}\right]+ \frac{T^2_{\scalebox{.55}{{\textsc{high}}}}N_c^2}{8}-\upkappa_0\,,&
\label{3.2}
\end{align}
where $T_{\scalebox{.55}{{\textsc{high}}}}$ is a sufficiently large temperature implying proximity to the ultraviolet fixed point. The temperature-dependent part, which is independent of the ultraviolet cutoff $\varepsilon$ of $\upkappa_\varepsilon(T_{\scalebox{.55}{{\textsc{high}}}})$ approaches $\upkappa_{\textrm{SYM}}(T_{\scalebox{.55}{{\textsc{high}}}})$, $u_{\scalebox{.55}{{\textsc{high}}}}=u_0(T_{\scalebox{.55}{{\textsc{high}}}})$, and $\upkappa_0$ is a constant to be subtracted to ensure that $\upkappa(T_{\scalebox{.59}{{\textsc{min}}}})=0$, where $T_{\scalebox{.59}{{\textsc{min}}}}$ is the lowest temperature considered in the numerical calculations ($T_{\scalebox{.59}{{\textsc{min}}}} \sim 10$ MeV) \cite{Finazzo:2014cna}.
Analogously, for evaluating the relaxation time for the shear relaxation   $\tau_\pi$ in Eq. \eqref{tt}, one can define 
\begin{align}
\tau_\pi&=\frac{1}{2\eta} \left(\Omega-\upkappa+T\frac{d\upkappa}{dT}\right),\label{3.4}
\end{align}
and 
\begin{align}
\Omega_\varepsilon&:=\lim_{\substack{{q}\to0\\\upomega\to0}}\frac{\partial^2}{\partial \upomega^2}\textsc{F}(\upomega,q;\varepsilon) =\frac{L^3}{2\varepsilon^2}+g_{xx}^{3/2}(u_0)\int_\varepsilon^{u_0} du \left(\frac{g_{xx}^{3/2}(u_0)}{\sqrt{-g^{\scalebox{.65}{(0)}}}g^{uu}(u)} -\frac{\sqrt{-g^{\scalebox{.65}{(0)}}}g^{tt}(u)}{g_{xx}^{3/2}(u_0)}\right),
\label{3.5}
\end{align}
Now the UV finite expression can be evaluated as 
\begin{align}
\Omega(T)&=\lim_{\varepsilon\rightarrow 0}\left(\Omega_\varepsilon(T) -\Omega_\varepsilon(T_{\scalebox{.55}{{\textsc{high}}}})\right) +\Omega_{\textrm{SYM}}(T_{\scalebox{.55}{{\textsc{high}}}})-\Omega_0\nonumber\\
&\approxeq \lim_{\varepsilon\rightarrow 0}\left(g_{xx}^{3/2}(u_0)\int^{u_0(T)}_\varepsilon du\left[\frac{g_{xx}^{3/2}(u_0)}{\sqrt{-g^{\scalebox{.65}{(0)}}}g^{uu}(u)}-\frac{\sqrt{-g^{\scalebox{.65}{(0)}}}g^{tt}(u)}{g_{xx}^{3/2}(u_0)}\right]\Biggr|_{u_0(T)} \right.\nonumber\\
&\left. \qquad -g_{xx}^{3/2}(u_{\scalebox{.55}{{\textsc{high}}}})\int^{u_{\scalebox{.55}{{\textsc{high}}}}}_\varepsilon du\left[\frac{g_{xx}^{3/2}(u_{\scalebox{.55}{{\textsc{high}}}})}{\sqrt{-g^{\scalebox{.65}{(0)}}}g^{uu}(u)}- \frac{\sqrt{-g^{\scalebox{.65}{(0)}}}g^{tt}(u)}{g_{xx}^{3/2}(u_{\scalebox{.55}{{\textsc{high}}}})} \right]\biggr|_{u_{\scalebox{.55}{{\textsc{high}}}}}\right) +\frac{T_{\scalebox{.55}{{\textsc{high}}}}^2N_c^2}{8}[1-\ln 2],
\label{3.6}
\end{align}
up to a finite constant that must be deducted to make certain that $\tau_\pi(T_{\scalebox{.59}{{\textsc{min}}}}) \eta(T_{\scalebox{.59}{{\textsc{min}}}}) = 0$. These integrals can be evaluated for $\varepsilon = 0.02$ \cite{Finazzo:2014cna}.

For the conformal super-Yang--Mills plasma, some of the following transport coefficients
\begin{subequations}
\beq
\tau_\pi&=&\frac{2-\ln 2}{2\pi T},\label{upt}\\
\uplambda_1&=&\frac{N_c^2 T^2}{16},\\
\uplambda_2 &=& -\frac{\ln 2}{8}N_c^2 T^2,\\
\uplambda_3&=& 2\upkappa-T\frac{d\upkappa}{dT},\\
\upkappa &=&\frac{T^2N_c^2}{8},\label{preview2}
\eeq
\end{subequations}
 carry corrections by quantum gravity, encoded into the running parameter $\upgamma$, 
where $T$ denotes the temperature of the QCD system.
Such corrections, introduced by the functional measure, can be taken into account when Eq. (\ref{gr1}) is promoted to 
\beq\label{gr11}
\!\!\!\!\!\!\!\!\!\!\!\!\!\!\!(\Uppsi^{\scalebox{.59}{\textsc{eff}}})^{\mu\nu} &=& -\eta^{\scalebox{.59}{\textsc{eff}}} \upsigma^{\mu\nu}- \eta^{\scalebox{.59}{\textsc{eff}}}\tau_\pi^{\scalebox{.59}{\textsc{eff}}} \left[  u_\rho\nabla^\rho {}^\llcorner\upsigma^{\mu\nu}{}^\lrcorner 
 + \frac 12 \upsigma^{\mu\nu}
    (\nabla \cdot \vec{u}) \right]  
  + \upkappa^{\scalebox{.59}{\textsc{eff}}}\left[R^{\llcorner\mu\nu\lrcorner}-2 u_\alpha R^{\alpha\llcorner\mu\nu\lrcorner\beta} 
      u_\beta\right]\nonumber\\
  &&+ \frac{\uplambda_1^{\scalebox{.59}{\textsc{eff}}}}{\left(\eta^{\scalebox{.59}{\textsc{eff}}}\right)^2} {\upsigma^{\llcorner\mu}}_\rho \upsigma^{\nu\lrcorner\rho}
  - \frac{\uplambda_2^{\scalebox{.59}{\textsc{eff}}}}{\eta^{\scalebox{.59}{\textsc{eff}}}} {\upsigma^{\llcorner\mu}}_\rho \Omega^{\nu\lrcorner\rho}
  + \uplambda_3^{\scalebox{.59}{\textsc{eff}}} {\Omega^{\llcorner\mu}}_\rho \Omega^{\nu\lrcorner\rho}+2 \left(\kappa^{*}\right)^{\scalebox{.59}{\textsc{eff}}} \,u_\rho u_\sigma {R}^{\rho \lrcorner \mu\nu \llcorner \sigma}\nonumber\\
  && -\eta^{\scalebox{.59}{\textsc{eff}}}\tau_\pi^{\scalebox{.59}{\textsc{eff}}} \left(\frac{1}{3}-c_s^2\right) \, \sigma^{\mu\nu}\nabla_\rho u^\rho+\uplambda_4 (\nabla^{\lrcorner\mu}\ln s)(\nabla^{\nu\llcorner} \ln s).
\eeq
Ref. \cite{Kuntz:2022kcw} showed that the speed of sound, appearing in the penultimate term in Eq. (\ref{gr11}), remains invariant under quantum gravity corrections. 
The coefficients carrying quantum corrections in the dissipative part of the energy-momentum tensor in Eq. (\ref{gr11}) can be related to the ones in Eq. (\ref{gr1}), which do not have quantum corrections, as:
\begin{subequations}
\beq
\tau_\pi^{\scalebox{.59}{\textsc{eff}}}&=&\tau_\pi+12\frac{(2-\ln 2)}{\pi T}\upgamma^2,\label{upt1}\\
\uplambda_1^{\scalebox{.59}{\textsc{eff}}}&=&\uplambda_1+\frac{13 N_c^2 T^2}{4}(13+\upgamma^2)\upgamma^2,\\
\uplambda_2^{\scalebox{.59}{\textsc{eff}}} &=& \uplambda_2 -\upgamma^2(6+\upgamma^2)\frac{\ln 2}{4}N_c^2 T^2,\\
\uplambda_3^{\scalebox{.59}{\textsc{eff}}}&=& 2\upkappa^{\scalebox{.59}{\textsc{eff}}}-T\frac{d\upkappa^{\scalebox{.59}{\textsc{eff}}}}{dT} = \uplambda_3,\\
\upkappa^{\scalebox{.59}{\textsc{eff}}} &=&\upkappa+\frac{3T^2N_c^2}{4}\upgamma^2.\label{preview31}
\eeq
\end{subequations}
Among the effective response and transport coefficients (\ref{upt1}) -- (\ref{preview31}), only $\uplambda_3$ does not receive any correction. 
The temperature scale is chosen to match the minimum speed of sound computed holographically with that found on the lattice results for (2 + 1)-flavor QCD \cite{katz1}. Ref. \cite{Finazzo:2014cna} obtained the pseudo-critical temperature for chiral crossover transition $T_c = 143.8$ MeV, indicating a crossover phase transition from the QGP to hadronic matter. It proposed the fit of the quantity 
$\frac{\tau_\pi \eta}{T^2}$ as a function of $\frac{T}{T_{c}}$, given by 
\beq\label{tpit2}
\frac{\tau_\pi \eta}{T^2} \!=\! \frac{0.2664}{1\!+\!\exp\left[2.029\left(0.7413\!-\!\frac{T}{T_{c}}\right)\right]\!+\!\exp\left[-0.1717\left(10.76\!+\!\frac{T}{T_{c}}\right)\right]\!+\!\exp\left[9.763\left(1.074\!-\!\frac{T}{T_{c}}\right)\right]}. \label{fit}
\eeq  
Therefore the running parameter $\upgamma$, driving corrections due to a functional measure encoding quantum gravity effects, can be realized as a temperature-dependent parameter. 
Data obtained from (2+1)-flavor lattice QCD to the pressure and the relaxation time can be compared to the ones whose quantum gravity corrections have been estimated in Secs. \ref{s3} and \ref{s4}, for temperatures in the range $130$ MeV $\lesssim T\lesssim 450$ MeV, where an entirely hadronic description is not appropriate enough. Besides, this range of temperature is not sufficiently high to ensure a straightforward formulation through perturbative aspects of QCD.

\begin{figure}[H]\begin{center}
\includegraphics[scale=0.75]{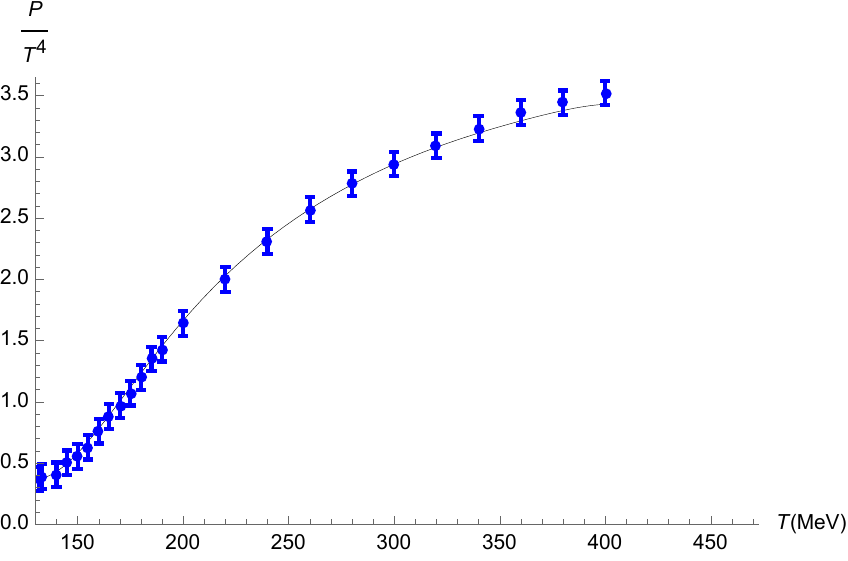}
\caption{\footnotesize Plot of $P/T^4$ as a function of $T$, for the bottom-up holographic model, indicated by the black curve. The results for the  lattice (2 + 1)-flavor QCD are plotted as blue points \cite{katz1}.}
\label{gmm100}\end{center}
\end{figure}

We first analyze the behavior of the pressure as a function of the temperature, depicted in Fig. \ref{gmm100}.
The function $P/T^4$ can be interpolated as a polynomial function of the temperature, for the bottom-up holographic model, as
\beq
\frac{P}{T^4} &=& -5.8077\times 10^{-16}\, T^7+ 1.1725\times 10^{-12}\, T^6 - 1.0057\times 10^{-9} \,T^5 +  4.7444\times 10^{-7}\, T^4  
\nonumber\\&&- 1.3266\times 10^{-4}\, T^3 + 2.1860\times 10^{-2}\, T^2 -1.9328\, T + 70.185, \eeq
within a 0.01\% root-mean-square deviation.

\begin{figure}[H]\begin{center}
\includegraphics[scale=0.75]{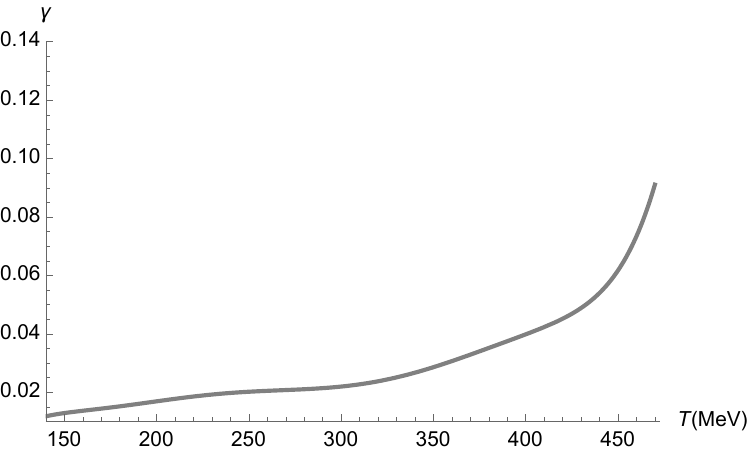}
\caption{\footnotesize Range of the running parameter $\upgamma$ (GeV$^4$) as a function of the temperature (MeV), using the numerical data  corresponding to Fig. \ref{gmm1}.}
\label{gmm355}\end{center}
\end{figure}

Using Eq. (\ref{peff}) yields
\beq\label{pt40}
\left|P^{\scalebox{.59}{\textsc{eff}}}\right| = P\sqrt{1+\frac{4\upgamma^2}{P^2}}.
\eeq One can therefore employ the data in Fig. \ref{gmm100} to obtain the running parameter $\upgamma$, as a function of the temperature, in Fig. \ref{gmm355}. In fact, the QGP carries quantum corrections due to the functional measure. Hence the left-hand side of Eq. (\ref{tpit2}), reading off the relationship between the relaxation time and the temperature, regards
the quantity $\frac{\tau_\pi^{\scalebox{.59}{\textsc{eff}}} \eta}{T^2}$. In its turn,  Eq. \eqref{upt1} relates the effective relaxation time, containing quantum corrections, to the relaxation time without quantum corrections. The left-hand side of Eq. (\ref{tpit2}) then yields
\beq
\frac{\tau_\pi \eta}{T^2}+12\frac{(2-\ln 2) \eta}{\pi T^3}\upgamma^2.
\eeq
From it, we can use Eq. (\ref{upt}) and solve Eq. (\ref{tpit2}) for the running parameter $\upgamma$ as a function of the temperature.
\noindent Eq. (\ref{upt1}) yields an effective relaxation time $\tau_\pi^{\scalebox{.59}{\textsc{eff}}}$ encoding quantum gravity effects, which can be related to the standard relaxation time $\tau_\pi$ in (\ref{upt}) without quantum corrections. 
To compute the shear relaxation coefficient, $\eta \tau_\pi/T^2$, one can alternatively use Eq. (15) of Ref. \cite{Moore:2010bu} and Eq. (48) of Ref. \cite{Czajka:2017bod}, namely
\beq
\eta \tau_\pi = -\frac12\lim_{\substack{q\rightarrow 0\\
 \upomega\to0}} \frac{\partial^2}{\partial\upomega^2} \mathsf{G}_R^{xy,xy}(\upomega,\bmq) + \frac12\lim_{\substack{q\rightarrow 0\\
 \upomega\to0}} \frac{\partial^2}{\partial q^2}\mathsf{G}_R^{xy,xy}(\upomega,\bmq).
\label{eq:viscosity_kubo1}
\eeq
 \begin{figure}[H]\begin{center}
\includegraphics[scale=0.75]{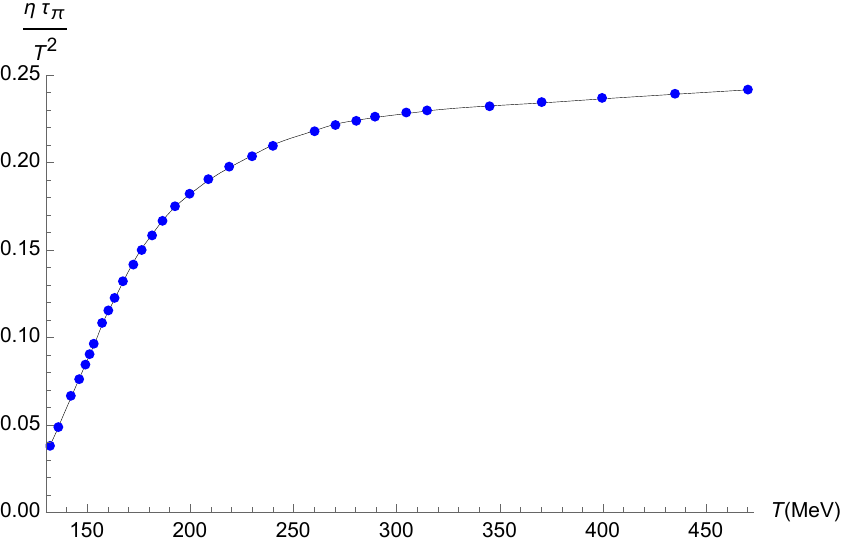}
\caption{\footnotesize Plot of $\tau_\pi \eta/T^2$ as a function of $T$ for the bottom-up holographic model. The blue points regard numerical data, whereas the black line corresponds to the fit in Eq. (\ref{fit}).}
\label{gmm3551}\end{center}
\end{figure}
\noindent Fig. \ref{gmm3551}  can be interpolated as a polynomial function of the temperature, for the bottom-up holographic model, as
\beq
\frac{\tau_\pi \eta}{T^2}&=& -3.3890 \times 10^{-17}\, T^7+ 7.1330\times 10^{-14}\, T^6 
 - 6.2508\times 10^{-11}\,T^5 +  2.9417\times 10^{-8}\, T^4 
\nonumber\\&& - 7.9647\times 10^{-6}\, T^3 + 1.2219\times 10^{-3}\,T^2 - 9.4946\times 10^{-2} T +
2.8185, \eeq
\noindent within a 0.01\% root-mean-square deviation.

\begin{figure}[H]\begin{center}
\includegraphics[scale=0.75]{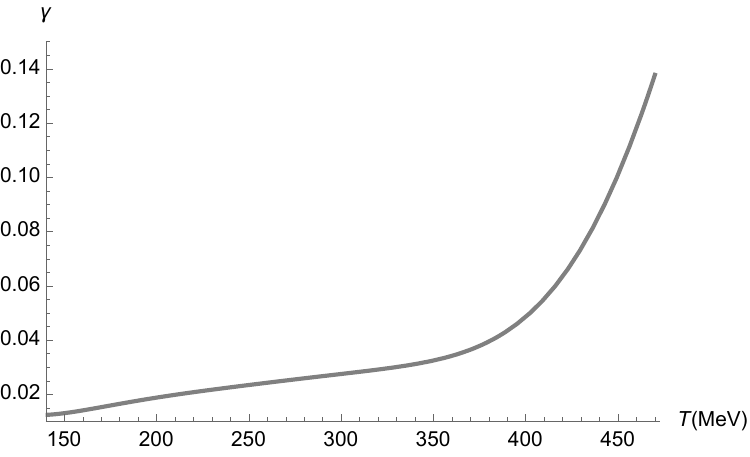}
\caption{\footnotesize Range of the running parameter $\upgamma$ (GeV$^4$) as a function of the temperature $T$ (MeV), using the numerical data  corresponding to Fig. \ref{gmm} and Eq. (\ref{upt}).}
\label{gmm4}\end{center}
\end{figure}
\noindent For the running parameter $\upgamma = \upgamma(\Lambda)$,  an energy scale $\Lambda \sim$ 3.0 TeV can be adopted in Eq.~\eqref{RGE}, as experimental results involving the QGP can arise in Pb+Pb and p-Pb collisions at $\sqrt{s_{\textsc{NN}}}$ = 2.76 and 5.02 TeV, at the ALICE experiment in the LHC.

\section{Bounding the parameter $\upgamma$ from experimental data from LHC and RHIC}
 \label{rhic}
 
\begin{figure}[H]\begin{center}
\includegraphics[scale=0.75]{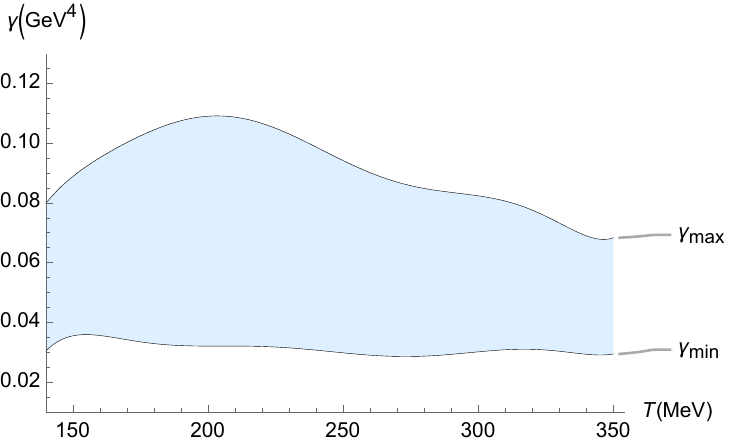}
\caption{\footnotesize Experimental bound on $\upgamma$ (GeV$^4$) as a function of the QGP temperature (MeV), using the experimental range of $\upzeta/s$ determined  by the JETSCAPE Bayesian model   \cite{JETSCAPE:2020shq,JETSCAPE:2022cob}.}
\label{gmm}\end{center}
\end{figure}

A reliable bound on the parameter $\upgamma$, which drives the quantum gravity  corrections due to the functional measure, can be dictated from  experimental data in LHC and RHIC, regarding the $\upzeta/s$ ratio of the QGP.  It can be compared with the ones whose quantum corrections have been  predicted in Secs. \ref{s3} and \ref{s4}. 
The values of transport coefficients of QGP have been precisely determined in heavy-ion collision experiments, for temperatures running in the range $150$ MeV $\lesssim T\lesssim 350$ MeV.  The lower limit in this range, at least for zero
baryon density or baryon chemical potential, approximately corresponds to the pseudo-critical temperature of a smooth crossover between the confined and the deconfined phase \cite{daRocha:2021xwq,Braga:2024nnj}. To determine a trustworthy bound on $\upgamma$,  Eq. (\ref{bul0}) can be employed, denoting \cite{Kuntz:2022kcw}
\beq\label{twt}
\left|\upzeta^{\scalebox{.59}{\textsc{eff}}}\right| = \upzeta\left(1+\frac{16\upgamma^2}{\upzeta^2}\right)^{1/2}.
\eeq
One can split the analyses of up-to-date experimental estimates for the QGP bulk viscosity into five independent parts. The first one takes into account the JETSCAPE Bayesian model  \cite{JETSCAPE:2020shq,JETSCAPE:2022cob}. 
Bayesian inference is employed to obtain probabilistic
constraints for $\upzeta/s$ from experimental and theoretical uncertainties. Bayesian Model Averaging accounts for the transition from a hydrodynamical fluid describing the QGP to hadronic transport in the ending evolution stage, yielding a reliable phenomenological constraint range for $\upzeta/s$ \cite{Gardim:2020mmy}. 
An experimental bound $\upgamma_{\text{min}} \lesssim \upgamma \lesssim \upgamma_{\text{max}}$, for the parameter carrying quantum gravity  corrections generated by a functional measure, is depicted in Fig. \ref{gmm} as a function of the QGP temperature.
\noindent One can interpolate the QGP temperature-dependent lower and upper bounds for $\upgamma$ in Fig. \ref{gmm}, using the polynomials:
\begin{subequations}
\beq\label{jetscape125}
\upgamma_{\scalebox{.59}{{\textsc{min}}}}(T)&=&4.47201\times 10^{-16} \,T^7-7.82417\times 10^{-13} \,T^6+5.78159\times 10^{-10}
   \,T^5-2.33781\times 10^{-7} \,T^4\nonumber\\
   &&\qquad\,+5.58432\times 10^{-5}
  \, T^3-7.87737 \times 10^{-2}\, T^2+0.60739\,T-19.7078,\\
  \upgamma_{\textsc{max}}(T)&=&6.01245\times 10^{-16} \,T^7-1.00634\times 10^{-12} \,T^6+7.08114\times 10^{-10}\,
   T^5-2.71365\times 10^{-7} \,T^4\nonumber\\
   &&\qquad\,+6.11704 \times 10^{-5} 
  \, T^3-8.12106\times 10^{-3} \, T^2+0.58963\, T-18.0404,\label{dukee25}
   \eeq\end{subequations}
within $10^{-3}\%$ interpolation error.

\textcolor{black}{The second part of the analysis consists of considering the experimental data of $\upzeta/s$ of the QGP, outlined by the Duke group \cite{Bernhard:2019bmu}. Their results present very precise estimates for the experimental value of  $\upzeta/s$ for the QGP.  It includes quantitative uncertainties from Bayesian parameter estimation protocols involving the analysis of a dynamical collision model and  experimental data. 
This time, the QGP temperature-dependent lower and upper bounds on $\upgamma$ are depicted in Fig. \ref{gmm1}.
\begin{figure}[H]\begin{center}
\includegraphics[scale=0.75]{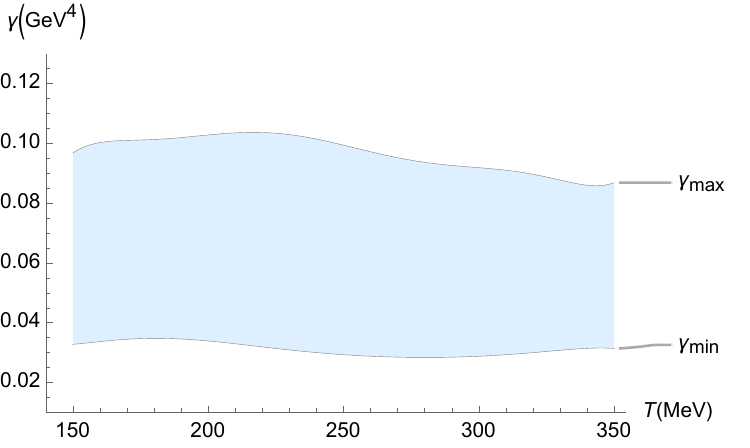}
\caption{\footnotesize Experimental bound on $\upgamma$ (GeV$^4$) as a function of the QGP temperature (MeV), using the experimental range of $\upzeta/s$ determined  by the analysis of the Duke group \cite{Bernhard:2019bmu}.}
\label{gmm1}\end{center}
\end{figure}
\noindent Such lower and upper bounds on $\upgamma$ in Fig. \ref{gmm1} can be respectively interpolated by 
\begin{subequations}
\beq\label{duke125}
\upgamma_{\scalebox{.59}{{\textsc{min}}}}(T)&=&-8.78999\times 10^{-17} \,T^7+1.54981\times 10^{-13}\, T^6-1.15801\times 10^{-10}
   \,T^5+4.74402\times 10^{-8} \,T^4\nonumber\\
   &&\qquad\,-1.14777\times 10^{-5}
   \,T^3+1.63481 \times 10^{-2}\, T^2-0.126542\,T+4.13012,\\
  \upgamma_{\textsc{max}}(T)&=&6.07662\times 10^{-16} T^7-1.06355\times 10^{-12} T^6+7.86621\times 10^{-10}
   T^5-3.18494\times 10^{-7} T^4\nonumber\\
   &&\qquad\,+7.62074 \times 10^{-5} 
   T^3-1.07744\times 10^{-2}  T^2+0.83363\, T-27.1427,\label{dukee25}
   \eeq\end{subequations}
within $10^{-3}\%$ interpolation error. }

\textcolor{black}{The up-to-date experimental results of $\upzeta/s$, implemented by the  Jyv\"askyl\"a-Helsinki-Munich group \cite{Parkkila:2021yha}, can be now used. 
They refer, in particular, to $\upzeta/s$ of the QGP in relativistic heavy-ion collisions, which are quantified through an improved global Bayesian analysis using the LHC Pb-Pb
 experimental data. 
The lower and upper bounds  for $\upgamma$, as a function of the QGP temperature, are illustrated in Fig. \ref{gmm35}. 
\begin{figure}[H]\begin{center}
\includegraphics[scale=0.75]{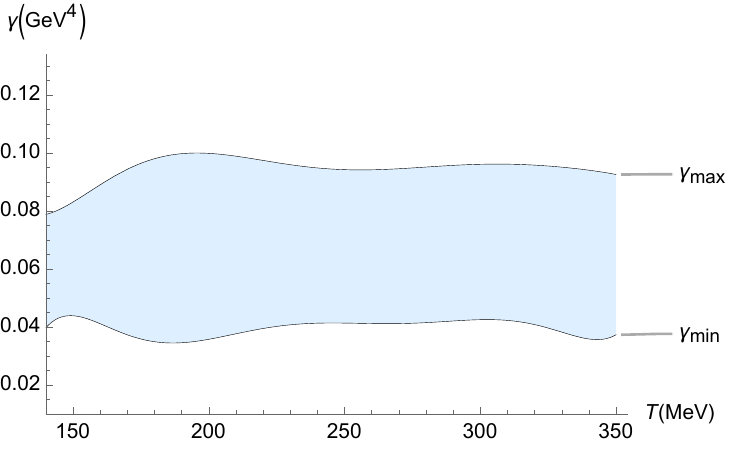}
\caption{\footnotesize \footnotesize Experimental bound on $\upgamma$ (GeV$^4$) as a function of the QGP temperature (MeV), using the experimental range of $\upzeta/s$ determined  by the Jyv\"askyl\"a-Helsinki-Munich group \cite{Parkkila:2021yha}.}
\label{gmm35}\end{center}
\end{figure}
\noindent The upper and lower bounds for $\upgamma$ in Fig. \ref{gmm35} can be respectively interpolated by the polynomials
\begin{subequations}
\beq\label{dukee1}
\upgamma_{\scalebox{.59}{{\textsc{min}}}}(T)&=&8.15869\times 10^{-16} \,T^7-1.42094\times 10^{-12}\, T^6+1.04585\times 10^{-9}
 \,  T^5-4.21334\times 10^{-7}\, T^4\nonumber\\
   &&\qquad\,+1.00242\times 10^{-4}
   T^3-0.014069\, T^2+1.07749\,T-34.6634,\\
  \upgamma_{\textsc{max}}(T)&=&-1.96075\times 10^{-16} \,T^7+3.64874\times 10^{-13} \,T^6-2.86110\times 10^{-10}\,
   T^5+1.22239\times 10^{-7} \,T^4\nonumber\\
   &&\qquad\,-3.06481 \times 10^{-5} 
  \, T^3+4.49533\times 10^{-3} \, T^2-0.35592\, T+11.7849,\label{dukee2}
   \eeq\end{subequations}
within $10^{-3}\%$ interpolation error. These results update the ones in Ref. \cite{Kuntz:2022kcw}. }

\textcolor{black}{The experimental data of the $\upzeta/s$ ratio for the QGP at LHC was analyzed by the MIT-Utrecht-Gen\`eve group, employing  the \textsc{Trajectum} setup, regarding an improved global Bayesian survey of the LHC Pb-Pb data at $\sqrt{s_{NN}} =  2.76$ and 5.02 TeV. Ref. \cite{Nijs:2022rme} demonstrated a non-negligible effect of the QGP $\upzeta/s$ in heavy-ion collision observables. This analysis takes  into account the measurements of higher-order harmonics in the hydrodynamical fluid flow and also in the flow fluctuation observables,  as inputs in the Bayesian analysis. 
The QGP temperature-dependent bound on $\upgamma$ is represented in Fig. \ref{gmmtra}.
\begin{figure}[H]\begin{center}
\includegraphics[scale=0.75]{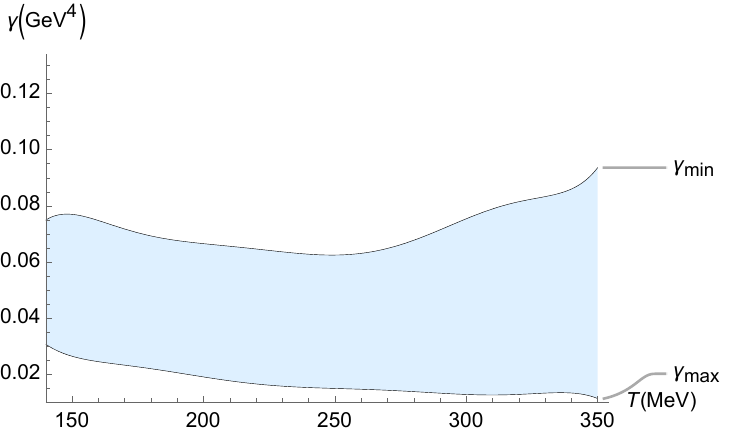}
\caption{\footnotesize \footnotesize \footnotesize Experimental bound on $\upgamma$ (GeV$^4$) as a function of the QGP temperature (MeV), using the experimental range of $\upzeta/s$ determined  from the analysis of experimental data of the QGP at LHC by the MIT-Utrecht-Gen\`eve group using the \textsc{Trajectum}  framework \cite{Nijs:2022rme}.}
\label{gmmtra}\end{center}
\end{figure} \noindent The lower and upper values of the bound for $\upgamma$ in Fig. \ref{gmmtra} are respectively represented by the following polynomials:
\begin{subequations}
\beq\label{dukee1}
\upgamma_{\scalebox{.59}{{\textsc{min}}}}(T)&=&6.57587\times 10^{-16} \,T^7-1.11752\times 10^{-12} \,T^6+8.01756\times 10^{-10}\,
   T^5-3.14696\times 10^{-7} \,T^4\nonumber\\
   &&\qquad\,+7.29776\times 10^{-5}\,
   T^3-9.99738\times 10^{-3}\, T^2+0.748838\,T-23.5695,\\
  \upgamma_{\textsc{max}}(T)&=&-2.67183\times 10^{-16} \,T^7+4.58546\times 10^{-13}\, T^6-3.32588\times 10^{-10}\,
   T^5+1.32076\times 10^{-7}\, T^4\nonumber\\
   &&\qquad\,-3.09994 \times 10^{-5} \,
   T^3+4.299319\times 10^{-3} \, T^2-0.32639\, T+10.4997,\label{mitg}
   \eeq\end{subequations}
within $10^{-3}\%$ interpolation error. }

\begin{figure}[H]\begin{center}
\includegraphics[scale=0.75]{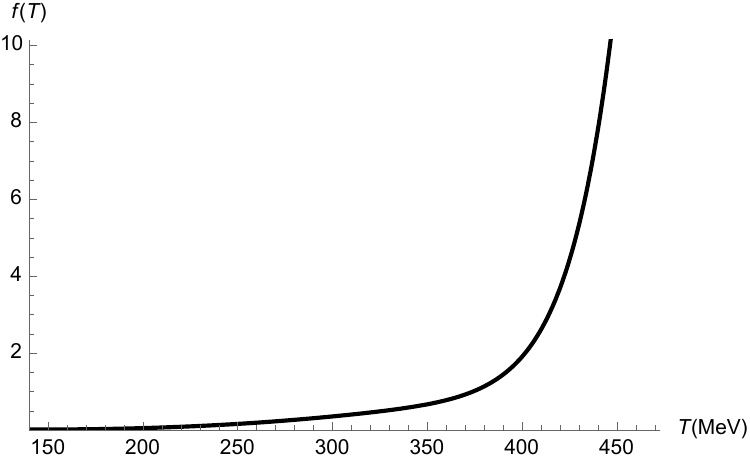}
\caption{\footnotesize Quantum correcting function $f(T)$, driving quantum corrections to the bulk viscosity-to-entropy density ratio in Eq. (\ref{twt}), as a function of the temperature (MeV).}
\label{ft25}\end{center}
\end{figure}

\textcolor{black}{Finally, the fifth independent part of this study uses 
the quark recombination model. The elliptic flows of $\phi$ and $\omega$ mesons, produced in Au+Au
collisions at $\sqrt{s_{NN}} = 200$ GeV, and in Pb+Pb collisions at $\sqrt{s_{NN}} = 2.76$ TeV are employed \cite{Yang:2022ixy}. Therefore, the temperature-dependent lower and upper bounds on the $\upgamma$ parameter are plotted in Fig. \ref{fshan}.
\begin{figure}[H]\begin{center}
\includegraphics[scale=0.75]{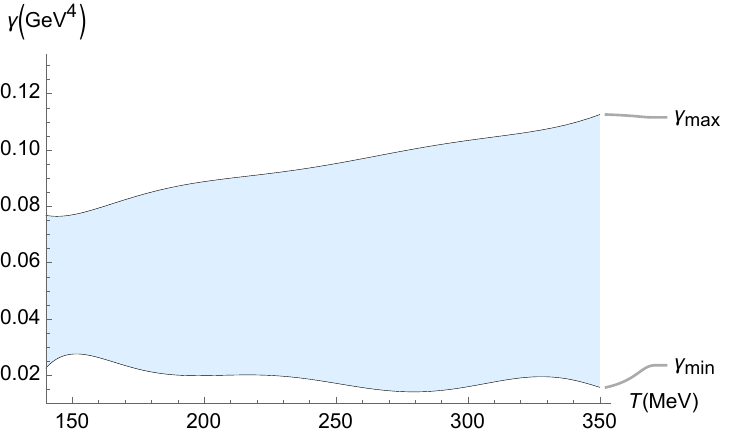}
\caption{\footnotesize \footnotesize \footnotesize Experimental bound on $\upgamma$ (GeV$^4$) as a function of the QGP temperature (MeV), using the experimental range of $\upzeta/s$ determined  by the Shanghai group \cite{Yang:2022ixy}.}
\label{fshan}\end{center}
\end{figure}
\noindent The QGP temperature-dependent lower and upper bounds for the running parameter $\upgamma$, depicted in Fig. \ref{fshan}, are interpolated by the polynomials
\begin{subequations}
\beq\label{dukee1s}
\upgamma_{\scalebox{.59}{{\textsc{min}}}}(T)&=&4.94397\times 10^{-16} T^7-8.86412\times 10^{-13} T^6+6.69641\times 10^{-10}
   T^5-2.76128\times 10^{-7} T^4\nonumber\\
   &&\qquad\,+6.70844\times 10^{-5}
   T^3-9.59746\times 10^{-3}\, T^2+0.748303\,T-24.4925,\\
  \upgamma_{\textsc{max}}(T)&=&-1.23113\times 10^{-16} T^7+2.29331\times 10^{-13} T^6-1.79668\times 10^{-10}
   T^5+7.66885\times 10^{-8} T^4\nonumber\\
   &&\qquad\,-1.92437 \times 10^{-5} 
   T^3+2.83603\times 10^{-3}  T^2-0.226875\, T+7.66334,\label{mshan}
   \eeq\end{subequations}
within $10^{-3}\%$ interpolation error.
}
A final remark is worth discussing, which involves the possibility of expressing the quantum corrected bulk viscosity-to-entropy density ratio as a deformation of the classical bulk viscosity-to-entropy density ratio, as 
 \beq\label{twt1}
\frac{\left|\upzeta^{\scalebox{.59}{\textsc{eff}}}\right|}{s^{\scalebox{.59}{\textsc{eff}}}} = \frac{\upzeta}{s}\left(1+f(T)\right),
\eeq
where $f(T)$ is some temperature-dependent function that regulates quantum corrections. Let us first remember that Eq. (\ref{seffs}) states that the entropy density does not carry any quantum corrections, as $
s^{\scalebox{.59}{\textsc{eff}}}  = s.$ 
Taking this into account, one can reasonably assume that  as quantum corrections are expected to be tiny, the running parameter $\upgamma$ satisfies $|\upgamma|\ll |\upzeta|$. Therefore, using Eq. (\ref{twt}), the bulk viscosity-to-entropy density ratio with quantum gravity corrections can be written in terms of the classical bulk viscosity-to-entropy density ratio, up to $\mathcal{O}(\upgamma^4/\upzeta^4)$, as
 \beq\label{twt1}
\frac{\left|\upzeta^{\scalebox{.59}{\textsc{eff}}}\right|}{s^{\scalebox{.59}{\textsc{eff}}}} = \frac{\left|\upzeta^{\scalebox{.59}{\textsc{eff}}}\right|}{s}
\approxeq \frac{\upzeta}{s}\left(1+\frac{8\upgamma^2}{\upzeta^2}\right).
\eeq
From Eq. (\ref{twt1}), one can realize that the required function $f(T)$ is given by 
\beq
f(T)=\frac{8\upgamma^2(T)}{\upzeta^2(T)}.
\eeq
Ref. \cite{Finazzo:2014cna} showed the  bulk viscosity-to-entropy density ratio $\upzeta/s$ as a function of the temperature $T$, for the bottom-up holographic model, with a resonance-like fitting function
\begin{equation}
\label{zetafit}
\frac{\zeta}{s}\left({T}\right) = \frac{0.01162}{\sqrt{\left(\frac{T}{T_c}-1.104\right)^2+0.05697}} - \frac{0.1081}{\frac{T^2}{T_c^2}+23.716},
\end{equation}
with $T_c=143.8$ MeV. The first term in Eq. \eqref{zetafit} implements a resonance-like peak, whereas the second term is responsible for a smooth background apart from the peak. 
We can substitute the temperature-dependent bulk viscosity  (\ref{zetafit}) on the very right-hand side of Eq. \eqref{twt1}, keeping in mind the expression for the entropy density $s=\pi^2 N_c^2 T^3/2$. The plot of the quantum correcting function $\upgamma=\upgamma(T)$ is shown in Fig. \ref{ft25}. 
\noindent 
Fig. \ref{ft25} indicates that the quantum correcting function $f(T)$, carrying quantum corrections to the bulk viscosity-to-entropy density ratio in Eq. (\ref{twt}), increases more sharply for $T\gtrsim 380$ MeV.  At the pseudo-critical temperature for the chiral crossover transition, $f(T_c)= 0.01578$, represeting a 1.57\% margin for quantum corrections to the classical $\zeta/s$ ratio.

\section{Discussions}
\label{s5}
{
\color{black}
In this paper, we used the gauge/gravity duality to determine transport coefficients of a non-conformal, strongly interacting non-Abelian plasma. This fluid displays a crossover transition similar to that found for the lattice calculations of QGP. We computed the contribution to transport and response coefficients, including the ones from the second-order expansion, due to the presence of a functional measure in quantum gravity. 
Comparison of the so-obtained pressure and relaxation time with existing data from lattice simulations led to the discovery of the temperature-dependence of the coupling constant $\upgamma$, which controls the functional-measure strength. In both cases, we found an upward trend of $\upgamma$ as the temperature increased. Even within the upper-temperature limit of the lattice data used, one finds $\upgamma\sim 10^{-1}$ GeV$^4$. \textcolor{black}{Since large-$N_c$ gauge theories can be employed to model QCD, the results obtained in this work can be applied in the scrutiny of several other aspects of the QGP.  Quantum gravity corrections implemented by a functional measure may be examined, at least in principle, in experiments   involving the QGP at the hydrodynamical regime, mainly the ones regarding the QGP transport coefficients. As shown in Eqs. (\ref{bul0}, \ref{bul}), only the $\upzeta/s$ ratio can detect quantum gravity effects, which are unseen by the $\eta/s$ ratio.  A relevant method was  implemented and illustrated in Figs. \ref{gmm} -- \ref{fshan}, showing the QGP temperature-dependent lower and upper bounds for the parameter $\upgamma$, carrying quantum gravity effects on $\upzeta/s$. It complements and refines the results already obtained in this work. 
The analysis relies on the up-to-date experimental results about the $\upzeta/s$ ratio measured for the QGP  \cite{JETSCAPE:2020shq,JETSCAPE:2022cob,Bernhard:2019bmu,Parkkila:2021yha,Nijs:2020roc,Nijs:2022rme,Yang:2022ixy}.} We conclude that the experimental range of the bulk viscosity-to-entropy density of the QGP, obtained by five different phenomenological analyses (JETSCAPE Bayesian model, Duke,  Jyväskylä-Helsinki-Munich, MIT-Utrecht-Gen\`eve, and Shanghai)  corroborate the existence of a non-vanishing renormalized parameter encoding the one-loop functional-measure quantum gravity correction, such that $10^{-2}$ GeV$^4 \lesssim \upgamma\lesssim 10^{-1}$ GeV$^4$. In this way, experimental data seem to favor the presence of a non-trivial functional measure. 
This suggests that a high-temperature scenario might be used to test the functional measure correction and ultimately serve as an experimental probe for quantum gravity.
}

\medbreak
\paragraph*{Declaration of competing interest.} The authors declare that they have no known competing financial interests or personal relationships that could have appeared to influence the work reported in this paper.
\paragraph*{Data Availability Statements:} The datasets generated during and/or analyzed during the current study are available from the corresponding author upon reasonable request.
\subsubsection*{Acknowledgments} 
\noindent IK thanks the National Council for Scientific and Technological Development -- 
CNPq (Grant No. 303283/2022-0 and Grant No. 401567/2023-0) for partial financial support. RdR~thanks to The S\~ao Paulo Research Foundation -- FAPESP
(Grants No. 2021/01089-1 and No. 2024/05676-7), and to the National Council for Scientific and Technological Development -- CNPq  (Grants No. 303742/2023-2 and No. 401567/2023-0),  and the Coordination for the Improvement of Higher Education Personnel (CAPES-PrInt  88887.897177/2023-00), for partial financial support.

\end{document}